Astronomy & Astrophysics

# How the super-Eddington regime affects black hole spin evolution in high-redshift galaxies

Warren Massonneau[1], Yohan Dubois[1], Marta Volonteri[1], and Ricarda S. Beckmann[1,2]

[1] Institut d'Astrophysique de Paris, CNRS, Sorbonne Université, UMR7095, 98bis bd Arago, 75014 Paris, France
e-mail: warren.massonneau@iap.fr
[2] Institute of Astronomy and Kavli Institute for Cosmology, University of Cambridge, Madingley Road, Cambridge CB3 0HA, UK

**ABSTRACT**

By performing three-dimensional hydrodynamical (3D MHD) simulations of a galaxy in an isolated dark matter halo, we are able to trace the evolution of the spin parameter $a$ of a black hole (BH) undergoing super-Eddington phases throughout its growth. This regime, suspected to be accompanied by powerful jet outflows, is expected to decrease the BH spin magnitude. We combined super-Eddington accretion with sub-Eddington phases (quasar and radio modes) and followed the BH spin evolution. Due to the low frequency of the super-Eddington episodes, relativistic jets in this regime are not able to decrease the magnitude of the spin effectively, as thin-disc accretion in the quasar mode inevitably increases the BH spin. The combination of super- and sub-Eddington accretion does not lead to a simple explicit expression for the spin evolution because of feedback from super-Eddington events. An analytical expression can be used to calculate the evolution for $a \lesssim 0.3$, assuming the super-Eddington feedback is consistently weak. Finally, BHs starting with a low spin magnitude are able to grow to the highest mass and if they initially start out as being misaligned with the galactic disc, they get a small boost of accretion via retrograde accretion.

## 1. Introduction

Quasars are believed to be powered by black holes (BHs) lying in the centres of galaxies, accreting gas and releasing their feedback in the form of powerful outflows. Astrophysical BHs are entirely characterised by two properties, namely, their mass and angular momentum, as they are thought to have zero charge (Blandford & Znajek 1977). A spinning BH of a mass, $M_{BH}$, has an angular momentum, $J_{BH}$, which is usually defined by its dimensionless spin parameter, $a = J_{BH}c/(GM_{BH})$, with $c$ being the speed of light and $G$ as the gravitational constant; it is described by the Kerr metric (Kerr 1963). The spin of a BH evolves as a result of its gas accretion or coalescence – and the relative importance of these effects varies along the mass scale (Volonteri et al. 2005, 2013; Berti & Volonteri 2008; Fanidakis et al. 2011; Barausse 2012; Dubois et al. 2014a; Sesana et al. 2014; Bustamante & Springel 2019). The spin of BHs changes how the feedback from the active galactic nuclei (AGN) proceeds, thus affecting its impact on galaxy properties, as it controls the efficiency of AGN feedback and the direction of jets in AGN (Dubois et al. 2014b; Beckmann et al. 2019, 2022; Cenci et al. 2021; Sala et al. 2021; Talbot et al. 2021, 2022; Huško et al. 2022; Massonneau et al. 2022).

Initial studies related to spin evolution started from a consideration of the accretion coming from cold discs and examining the consequences of the Kerr metric. Bardeen (1970) assumed that the angular momentum of the gas and the BH spin were always aligned, causing $a \to 1$ after a moderate amount of mass is accreted by the compact object. In such a cold disc scenario, because of the radiation emitted from the disc at high spins, Thorne (1974) discovered that the maximum spin a BH could reach is $a \simeq 0.998$. However, not all black holes are thought to have a cold gas disc at all times, and depending on the state of the accretion disc as well as the associated feedback modes, the spin can evolve very differently.

The question of how efficiently a BH accretes can be parameterised in relation to the Eddington limit ($L_{Edd} \equiv 4\pi GM_{BH}m_p c/\sigma_T$, with $m_p$ as the proton mass and $\sigma_T$ as the Thomson cross-section). When accretion rates are much smaller than the Eddington limit, the accretion is radiatively inefficient (or advection dominated) and accompanied by relativistic outflows, which may be launched via the Blandford-Znajek mechanism (Blandford & Znajek 1977). The process of powering a jet leads to a loss of angular momentum, meaning that the BH would spin down, namely, decrease its magnitude $|a|$. Early works based on general relativistic, magnetohydrodynamical (GRMHD) simulations (e.g. Gammie et al. 2004; De Villiers et al. 2005; Komissarov et al. 2007; McKinney & Blandford 2009) opened a window on many insights regarding the role of magnetic fields in accretion flows, especially related to the extraction of power from spinning BHs. Early studies predicted an equilibrium spin in the range of $a \simeq 0.9$–$0.94$ would be reached. Over the next decade, key developments came about in the area of accretion theory in relation to the importance of magnetically-dominated accretion flows. These findings have led to the magnetically arrested disc (MAD) model (Igumenshchev et al. 2003; Narayan et al. 2003), which demonstrates that BHs surrounded by such discs are able to produce jets whose energy exceeds the rest mass of the accreted gas (Tchekhovskoy et al. 2011). Works such as Tchekhovskoy et al. (2012), McKinney et al. (2012), and Narayan et al. (2022) have asserted that BHs in MADs may spin down more efficiently, with an equilibrium spin of $a \simeq 0.03$–$0.08$.

Accretion rates higher than the Eddington limit give a different disc structure than the typical thin disc at $L \sim L_{Edd}$. Different theoretical models have been developed in the past decades (see, e.g. a comprehensive review by Mayer & Bonoli 2019) and one of the most widely adopted to explain this critical state of accretion is the slim-disc model (Abramowicz et al. 1988; Sądowski 2009; Abramowicz & Fragile 2013). Gas flowing towards the BH proceeds via an optically and geometrically thick accretion disc, which traps the photons emitted from gas accretion and makes this process radiatively inefficient (e.g. Katz 1977; Begelman 1978; Ohsuga et al. 2005; Madau et al. 2014). Such super-Eddington accretion may be accompanied by powerful jets (Sądowski & Narayan 2015; Narayan et al. 2017), which could then have a significant impact on BH growth (Regan et al. 2019; Massonneau et al. 2022).

In the present study, we investigate, for the first time, the spin evolution of a BH undergoing both super- and sub-Eddington accretion in idealised hydrodynamical simulations of isolated galaxies, using the suite of initial conditions from Massonneau et al. (2022, hereafter Paper I). By extending the BH spin model from Dubois et al. (2021) in RAMSES (Teyssier 2002) to the super-Eddington regime, we explore the different phases of spin evolution when super-Eddington accretion is involved, as well as its impact on BH growth. We attempt to describe explicitly the evolution in order to predict the final value of $a$ after a given mass is accreted. The structure of the paper is as follows. We describe the BH spin models used for different accretion rates in Sect. 2, along with the setup of our simulations in Sect. 3. We show and discuss our simulations results in Sects. 4 and 5, finally presenting our conclusions in Sect. 6.

## 2. Impact of the disc structure on BH spin

Accretion from a gaseous disc orbiting a BH modifies both spin magnitude and direction. Following Shapiro (2005), we define the dimensionless spin-up parameter, $s$,

$$s \equiv \frac{\mathrm{d}a}{\mathrm{d}t}\frac{M_{\mathrm{BH}}}{\dot{M}_{\mathrm{acc}}} = \frac{\mathrm{d}a}{\mathrm{d}\ln M_{\mathrm{BH}}}(1 - \epsilon_{\mathrm{r}}), \quad (1)$$

where $a$ is the spin parameter, $M_{\mathrm{BH}}$ is the BH mass, $\epsilon_{\mathrm{r}}$ is the radiative efficiency, $\dot{M}_{\mathrm{acc}}$ corresponds to the accretion rate on the accretion disc, and the accretion rate on the BH is $\dot{M}_{\mathrm{BH}} = (1 - \epsilon_{\mathrm{r}})\dot{M}_{\mathrm{acc}}$. Here and in the following, we assume that a prograde BH is characterised by $a > 0$ and a retrograde one by $a < 0$. Furthermore, if $s$ and $a$ are of the same sign, then the BH spins up, namely, it increases its magnitude $|a|$, and vice versa: if $s$ and $a$ are of the opposite sign, the BH spins down, that is, it decreases its magnitude.

The spin-up parameter corresponding to the standard thin disc model, which we name $s_{\mathrm{qso}}$, is given by:

$$s_{\mathrm{qso}} = \widetilde{L}_{\mathrm{qso}} - 2a\widetilde{E}_{\mathrm{qso}}, \quad (2)$$

where $\widetilde{E}_{\mathrm{qso}} \equiv 1 - \epsilon_{\mathrm{r}}$ and $\widetilde{L}_{\mathrm{qso}}$ are respectively the energy and angular momentum of the innermost stable circular orbit (ISCO), $r_{\mathrm{ISCO}}$. Combining and integrating Eqs. (1) and (2), Bardeen (1970) obtained the following evolution law:

$$a_{\mathrm{fin}} = \frac{\sqrt{r_{\mathrm{ISCO,ini}}}}{3}\frac{M_{\mathrm{BH,ini}}}{M_{\mathrm{BH,fin}}}\left[4 - \sqrt{3r_{\mathrm{ISCO,ini}}\left(\frac{M_{\mathrm{BH,ini}}}{M_{\mathrm{BH,fin}}}\right)^2 - 2}\right]$$

$$\text{for } \frac{M_{\mathrm{BH,fin}}}{M_{\mathrm{BH,ini}}} < \sqrt{r_{\mathrm{ISCO,ini}}}, \quad (3)$$

$$= 1 \text{ otherwise.}$$

"ini" ("fin") are the initial (final) value of the quantity measured before (after) accretion respectively. This disc model has a positive spin-up parameter for all spin values up to $a = 0.998$ (Thorne 1974), meaning that counter- (co-)rotating accretion discs always spin the BH down (up).

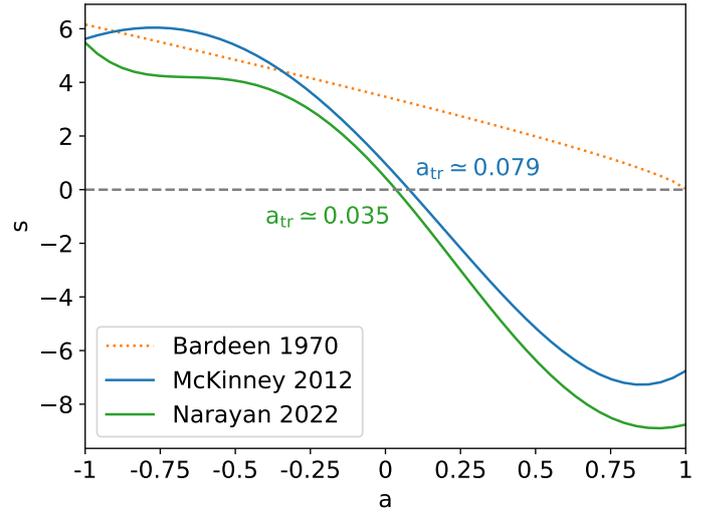

**Fig. 1.** Spin-up parameters $s$ for different disc models (thin disc in dotted orange, MAD in solid lines) as a function of the BH spin parameter $a$. Similarly to the thin disc, a MAD counter-rotating ($a < 0$) disc causes the BH to spin down. Co-rotating ($a > 0$) thin discs will spin the BH up, unlike the MAD ones (due to loss of angular momentum to the jets). A special MAD case for very low spins ($0 \leq a < a_{\mathrm{tr}}$) shows a weak spin-up, but is less than the spin-up from thin discs.

Thick discs, as well as other disc models such as the advection dominated accretion flow (ADAF, Narayan & Yi 1995; Manmoto et al. 1997; Narayan et al. 1997), or slim discs exhibiting a geometry that favours relativistic jet outflows, are able to collimate the jet because of this particular geometry. These jets are powered by energy extracted from the rotation of the BH, via the Blandford-Znajek mechanism, which, as a consequence, causes the spin magnitude $|a|$ to decrease. Recent GRMHD simulations studying spin evolution in MADs (e.g. McKinney et al. 2012; Narayan et al. 2022) have been able to demonstrate this spin-down effect. These accretion discs accumulate large-scale magnetic flux on the BH, until this flux becomes so strong that it chokes the gas infall. The excess magnetic field leaves and pushes parts of the disc away: the accretion flow enters a MAD state. Although the spin-up rates provided by McKinney et al. (2012) and Narayan et al. (2022) are for the sub-Eddington regime for thick discs, since the geometry of the discs are similar, and in absence of any spin-up rate prescription in the literature for the super-Eddington regime, we assume the same spin-up rates during the super-Eddington regime. The spin-up parameter we adopted, which we name $s_{\mathrm{MAD,MK12}}$, is taken from McKinney et al. (2012) and can be fitted with the following fourth-order polynomial (Dubois et al. 2021):

$$s_{\mathrm{MAD,MK12}} = 0.97 - 12.00a - 4.04a^2 + 5.81a^3 + 2.50a^4, \quad (4)$$

and, for comparison, the spin-up parameter from Narayan et al. (2022) fitted with a fifth-order polynomial is:

$$s_{\mathrm{MAD,N22}} = 0.45 - 12.53a - 7.8a^2 + 9.44a^3 + 5.71a^4 - 4.03a^5. \quad (5)$$

We show, in Fig. 1, the spin-up parameter $s$ for the standard thin disc and the ones found for MAD, as a function of $a$. In both McKinney et al. (2012) and Narayan et al. (2022), the authors find that non-spinning BH ($a = 0$) experiences a mild spin-up ($s > 0$). The weak spin-up is a result of the inflowing gas having non-zero angular momentum, which spins the BH up. Therefore, for certain spins $a \in [0, a_{tr}]$ with $a_{tr}$ as the transition from $s_{MAD} > 0$ to $s_{MAD} < 0$, these models predict a BH spinning up as a consequence of weak loss of angular momentum from jets, in comparison to gas accretion. Studies related to the spin-up parameter for super-Eddington discs have not yet been explored. This regime is associated with radiatively inefficient accretion, due to the photon trapping effect, meaning $\epsilon_r \ll 1$ for higher accretion rates. Therefore, in this work, we assume that for the super-Eddington regime, Eq. (1) can be approximated by

$$s_{sEdd} \simeq \frac{da}{d \ln M_{BH}}. \tag{6}$$

## 3. Simulation set-up

This paper presents a set of hydrodynamical simulations of an isolated galaxy in its host dark matter halo at $z = 4$, produced using the adaptive mesh refinement code RAMSES. The addition of the accretion and feedback phases in the super-Eddington regime to RAMSES are described in Paper I, however, we briefly recall their implementation here, along with the spin evolution in RAMSES. We then describe the galaxy formation and BH seeding processes to achieve super-Eddington accretion rates.

### 3.1. Technical details in RAMSES

In RAMSES, BHs are represented by a 'sink' particle that can transfer mass, momentum and energy from and to the gas. In our simulations, they are manually placed with a given initial mass, velocity, and spin, at a certain point in time. The mass of these BHs grows at a rate $\dot{M}_{BH} = (1 - \epsilon_r)\dot{M}_{BHL}$, with $\dot{M}_{BHL}$ the Bondi-Hoyle-Littleton (BHL) accretion rate (Hoyle & Lyttleton 1939; Bondi 1952):

$$\dot{M}_{BHL} = \frac{4\pi G^2 M_{BH}^2 \bar{\rho}}{(\bar{c}_s^2 + \bar{v}_{rel}^2)^{3/2}}, \tag{7}$$

where the averaged density $\bar{\rho}$, sound speed $\bar{c}_s$ and relative velocity between the BH and the gas $\bar{v}_{rel}$ are computed within $4\Delta x$ of the BH, using mass weighting and kernel weighting as specified in Paper I. The accretion rate is not capped at the Eddington limit (unless otherwise stated), only by the total mass contained in the kernel divided by the timestep. We note that there is always enough mass available for BH accretion and feedback processes, with mass conservation enforced at all times.

The BH accretion is parameterised using the Eddington limit $L_{Edd}$. For very low accretion rates, when $f_{Edd} \equiv L/L_{Edd} \leq 0.01$, accretion is radiatively inefficient and accompanied by relativistic outflows. The BH enters a so-called 'radio' mode and the energy injected as kinetic energy follows Sądowski et al. (2016), giving a total jet feedback of:

$$\dot{E}_{jet} = \eta_{jet} \dot{M}_{BH} c^2, \tag{8}$$

where $\eta_{jet} = 1.3 a^2 f_{MAD}^2$ is the efficiency factor of the kinetic feedback for a MAD taken from Tchekhovskoy (2015), and $0 \leq f_{MAD} \leq 1$ is the fraction of MAD strength (magnetic field saturation).

On the other hand, for higher but sub-Eddington accretion rates ($0.01 < f_{Edd} \leq 1$), the AGN is in the 'quasar' mode, corresponding to feedback coming from winds and radiation. In

this regime, thermal energy is released and the total feedback deposited is

$$\dot{E}_{thm} = \eta_{thm} \dot{M}_{BH} c^2, \tag{9}$$

with $\eta_{thm}$ as the spin dependant thermal wind efficiency of the disc.

With our current understanding of the super-Eddington regime ($f_{Edd} > 1$), super-Eddington AGN feedback includes both kinetic and radiative/thermal components. The super-Eddington implementation injects both thermal and kinetic energy simultaneously, with the total AGN feedback in this regime based on Sądowski et al. (2016)

$$\dot{E}_{sEdd} = (\eta_{jet} + 0.5\eta_{thm})\dot{M}_{BH} c^2. \tag{10}$$

Dubois et al. (2021) implemented routines to follow on-the-fly the evolution of the spin parameter $a$. In addition to spin evolution below the Eddington limit, we now use the spin-up parameter, $s_{sEdd} = s_{MAD,MK12}$, (Eq. (4)) for computing the BH spin at super-critical accretion rates, since relativistic jets may be present in the super-Eddington regime.

In most general cases, a misalignment between the accretion disc and the BH spin occurs. Due to the Lense-Thirring effect, it generates a torque which makes the disc precess around the axis of the BH spin. This creates a warped disc, with a certain radius which marks the transition between the equatorial inner disc aligned with the BH spin and the misaligned outer disc. Lense-Thirring precession results in the BH and disc angular momentum being aligned or anti-aligned with the total angular momentum of the BH+disc system. King et al. (2005) find a criterion for the anti-alignment of the BH with the disc:

$$\cos\theta < -\frac{J_d}{2J_{BH}}, \tag{11}$$

where $J_d$ is the disc angular momentum and $\theta$ corresponds to the angle between the disc and disc angular momentum. For instance, if $\cos\theta \geq 0$, both angular momentum always align; while for $\cos\theta < 0$, the anti-alignment occurs for small values of $J_d/J_{BH}$. Details regarding the implementation of this process are given in Dubois et al. (2014a).

### 3.2. Galaxy and BH

Super-Eddington accretion is sustained onto the BH via strong gas inflows, which are more likely to be found in gas-rich high-redshift environments. To this end, we set up our initial conditions to represent an isolated halo of mass $10^{11} M_\odot$ at redshift $z = 4$ with an NFW profile composed of 85 per cent of DM particles and 15 per cent of gas in hydrostatic equilibrium with small rotation, where the gas can radiatively cool down to temperatures as low as 10K. The simulations presented here are performed in a box of size 113 kpc, adaptively refined to a maximum resolution of 12 pc. Stars form following a Schmidt law with a gravo-turbulent model for star formation efficiency, in cells above a gas density threshold of $n_{SF} = 181$ H cm$^{-3}$, giving a stellar mass resolution of $M_* = 10^4 M_\odot$. Explosions from type II supernovae (SNe) are also included with release of mass (including metals), momentum, and energy (assuming $2 \times 10^{49}$ erg $M_\odot^{-1}$ of formed stars). For more details about the initial conditions and the modeled physics, we refer to Paper I.

As the DM halo starts to collapse, our isolated galaxy begins to form. With enough cycles of star formation and SN feedback events, the galaxy comes to a steady state at $t = 160$ Myr with a total stellar mass of $\sim 10^9 M_\odot$. It corresponds to the stellar mass

**Table 1.** Properties of the suite of simulations performed.

| Name | $M_{BH}$ $(M_\odot)$ | $a_{ini}$ $(x, y, z)$ | $f_{MAD}$ |
|---|---|---|---|
| **sEdd_0** (fid.) | $\mathbf{2 \times 10^6}$ | **(0,0,0)** | **0.5** |
| sEdd_$y$ | $2 \times 10^6$ | (0,0.7,0) | 0.5 |
| sEdd_$-z$ | $2 \times 10^6$ | (0,0,-0.7) | 0.5 |
| sEdd_$+z$ | $2 \times 10^6$ | (0,0,0.7) | 0.5 |
| sEdd_0.05_0 | $2 \times 10^6$ | (0,0,0) | 0.05 |
| sEdd_0.05_$y$ | $2 \times 10^6$ | (0,0.7,0) | 0.05 |
| sEdd_0.05_$-z$ | $2 \times 10^6$ | (0,0,-0.7) | 0.05 |
| sEdd_$+z$_LM | $10^6$ | (0,0,0.7) | 0.5 |

**Notes.** Showing from left to right: BH mass when super-Eddington is allowed ($M_{BH}$); initial BH spin ($a_{ini}$); and MADness fraction of the disc ($f_{MAD}$, scales with the jet efficiency as $\eta_{jet} \propto f_{MAD}^2$). Values in bold correspond to the fiducial simulation.

expected in a $10^{11} M_\odot$ dark matter halo (Moster et al. 2010). In addition, the star formation rate reaches a few $10 M_\odot \, \mathrm{yr}^{-1}$, which is also the expected order of magnitude for our target redshift $z = 4$ (Salmon et al. 2015). Despite the strong SN explosions, due to the high densities of gas available for star formation, the centre of the galaxy remains cold, dense, and compact.

At this point, a BH is added in the centre of the galaxy, with a given velocity similar to the gas surrounding the BH, in order to maximise the BHL rate (see Eq. (7)). We do not allow for super-Eddington accretion as soon as the BH is added, the very high accretion rates would produce extremely strong feedback that would sterilize the BH environment. We look for a smooth transition from a galaxy without a BH to a BH with potentially very strong feedback by enforcing a few cycles of Eddington-limited accretion, so that the evolution of the galaxy remains more continuous. After ~40 Myr, at $t = 206.4$ Myr, we let the BH grow at super-Eddington rates – and our analysis starts from this time.

In order to study the super-Eddington regime in an idealised setup, it is important that the galaxy has relaxed and stabilized from its initial conditions and that one has to avoid blasting the galaxy with a burst of feedback that would be unrealistic in a naturally-evolving galaxy in a cosmological setting. This intermediate sub-Eddington phase prevents such issues, and it would not be necessary in a full cosmological context. A more detailed discussion on these choices, as well as issues related to star formation and SN feedback, is given in Paper I.

## 4. Results

In this work, we use the suite of initial conditions from Paper I in the same halo with properties analog to a $10^{11} M_\odot$ dark matter halo at $z = 4$ (see Table 1 for more technical details). Similarly to the super-Eddington simulations from Paper I, and as described in the previous section, we add the BH in the centre of the galaxy at $t = 160$ Myr, where it evolves under Eddington-limited accretion until ~206.4 Myr. From this point onwards, we allow for super-Eddington accretion and feedback processes. All figures shown in this Section start when super-Eddington is allowed. We evolved both the spin magnitude and direction in these new simulations to highlight the effects of the super-Eddington regime on the subsequent development of the BH spin (Sect. 4.1), as well as its impact on BH growth (Sect. 4.2). We then see how we can predict the BH spin according to simulations and analytical modelling (Sect. 4.3). All the simulations presented are summarized in Table 1.

### 4.1. Different phases of spin evolution

In order to study the evolution of $a$, we use as our fiducial simulation an initially non-spinning BH (sEdd_0). We show, in Fig. 2, the spin evolution $a$ against the mass gained by the BH $\Delta M_{growth}$ in this simulation, after the first super-Eddington episode (at $t \sim 206.4$ Myr). Specific snapshots are indicated by the coloured markers and for each marker, a zoomed-in evolution of the spin is added in the respective coloured panels.

Over the course of the simulation, the prograde BH spin tends to generally increase, reaching $a \simeq 0.5$ after gaining more than $3 \times 10^6 M_\odot$. As explained in Sect. 2, when $a > a_{tr} = 0.079$, the quasar mode will spin-up the BH, whereas the super-Eddington and radio modes will spin down the BH. It is important to note that mass accreted in the radio mode does not significantly contribute to BH growth and consequently to varying $a$ throughout the entirety of the simulation. In our setup, the super-Eddington regime does not seem to be frequent enough or to last long enough to have a durable impact on the BH spin. A careful inspection of the evolution shown in the panels of Fig. 2 hints at the specific reasons for the spin evolution of our BH, and we detail them in the following.

Shortly after critical accretion is allowed (blue panel), the spin increases during both super- and sub-Eddington accretion phases, since $a < 0.079$ during this period, that is, below the spin-up and spin-down transition, $a_{tr}$, used in our model for the MAD state. Most of the BH mass is gained via super-Eddington accretion (yellow line, inset panels of Fig. 2), meaning that super-Eddington accretion is the main driver of the spin evolution at this time in the simulation. There are however a few events of accretion in the quasar regime (purple line) that do not significantly impact the mass gained, but increase $a$ at a faster rate (per unit of mass). This is caused by the spin-up parameter of the thin accretion disc being significantly larger than that of the MAD state at this low value of the BH spin (see Fig. 1 for small values of $a$).

At $t = 209.3$ Myr in the orange panel, the picture is slightly different, as $a \sim 0.3$ and super-Eddington accretion do not have a significant impact on the BH spin evolution anymore ($s_{sEdd} \approx 0$ at $a \sim a_{tr}$, see Fig. 1). This leads to the almost flat evolution of the spin during each super-Eddington episode. Sub-Eddington accretion events lead to a continuous spin increase (with most of the mass acquired in the quasar mode), going past $a_{tr}$. During this phase, the BH spin evolution is dominated by sub-Eddington accretion, while the mass growth remains dominated by the super-Eddington episodes.

From here on, the spin slowly increases with more mass accreted. Repetitive patterns, such as those seen in the green panel at $t = 220$ Myr, emerge: quasar accretion phases that significantly grow the BH spin are intertwined with a few super-Eddington episodes, which decrease the BH spin. Overall, the mass accreted in quasar mode is larger than the mass accreted in the super-Eddington regime and the BH spin increases.

We find that, over time, the frequency of super-Eddington episodes slowly decreases and the spin magnitude grows more steadily, as shown in the red panel at $t = 246$ Myr. Despite $|s_{sEdd}| > s_{qso}$ for $a \gtrsim 0.3$ (see Fig. 1), the scarce super-Eddington episodes cannot spin down the BH efficiently. The more powerful spin-driven jets in the super-critical regime lead to less frequent accretion in super-Eddington, and this regime does not majorly contribute to the BH mass anymore, leading to a less effective spin-down. During this time the main role of super-Eddington episodes is to delay the overall spin-up caused by the quasar mode.

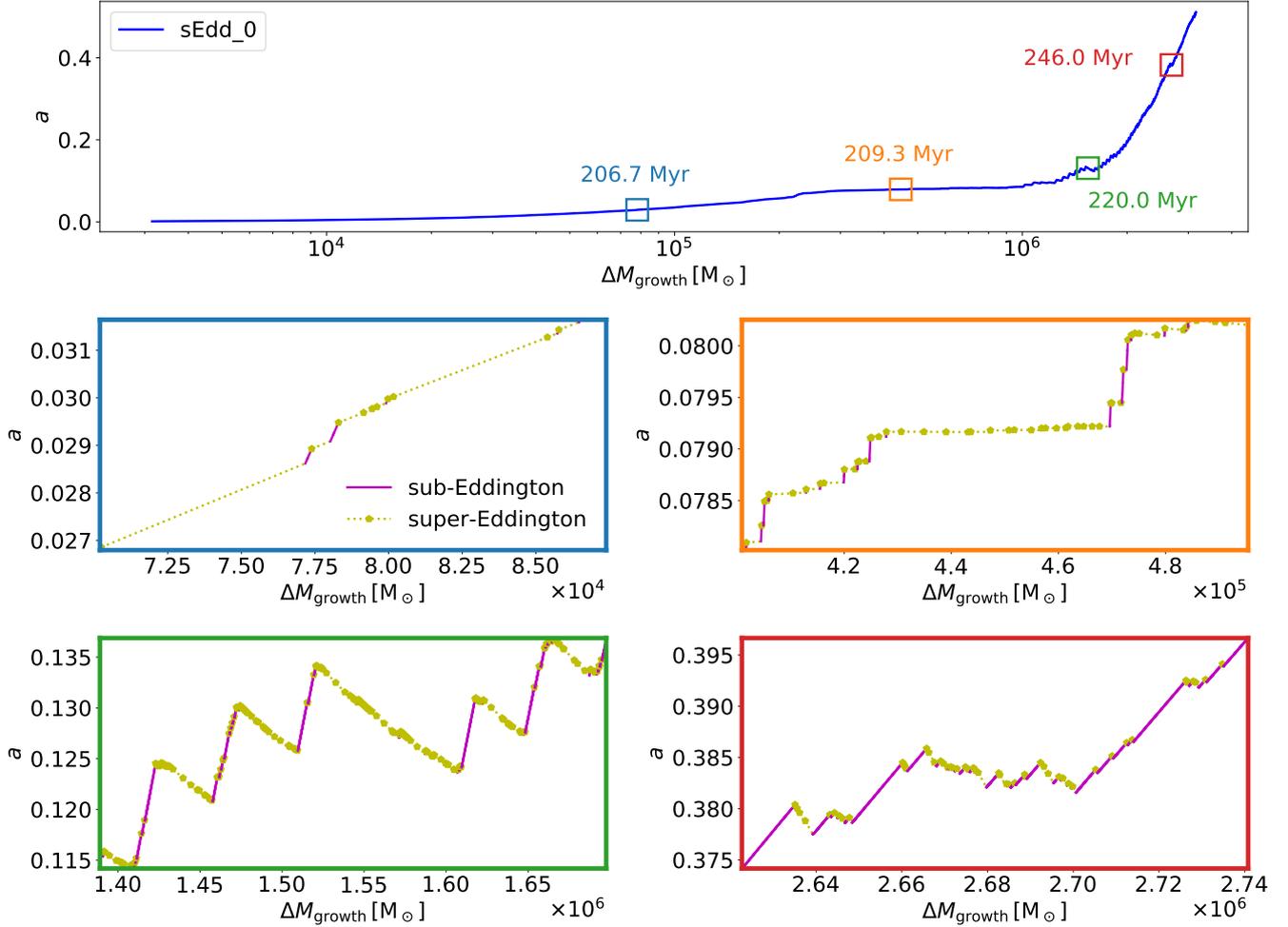

**Fig. 2.** Evolution of the spin parameter $a$ against the mass gained by the BH $\Delta M_{\mathrm{growth}}$ in sEdd_0, after the first super-Eddington episode (at $t \sim 206.4$ Myr). *Top*: spin parameter $a$ versus the cumulative BH mass growth $\Delta M_{\mathrm{growth}}$ for our fiducial simulation sEdd_0, after the first super-Eddington episode (at $t \sim 206.4$ Myr). Different phases of evolution are shown in the coloured panels. *Middle and bottom*: zoomed-in evolution of $a$ at different times where each colour of the plot frame corresponds to that in the *top panel*. Mass is either accreted above (dotted yellow) or below (solid purple) the Eddington limit; $a$ tends to increase with more mass gained by the BH and super-Eddington episodes are not able to effectively spin down the BH.

As seen in Paper I, more powerful super-Eddington episodes lead to very low accretion events in the radio mode ($f_{\mathrm{Edd}} < 0.01$). Due to their low accretion rates, these radio mode episodes neither significantly increase BH mass nor have any noticeable impact on the BH spin, and are therefore not visible on either panels.

If we were to compare the evolution of our BH shown here with an Eddington-limited run, the spin of our BH would generally increase more slowly at a given $\Delta M_{\mathrm{growth}}$. For instance, using Eq. (3), a BH constantly accreting in the quasar mode would reach $a \approx 0.5$ (from $a = 0$) after increasing its mass by a factor $\sim 1.19$ ($\Delta M_{\mathrm{growth}} \approx 4 \times 10^5\ M_\odot$). Our fiducial simulation has a BH reaching $a \simeq 0.5$ after accreting more than seven times this mass.

In our analysis, we find that our BH inevitably spins up, but the super-Eddington regime is able to delay this process. For low to moderate ($a \lesssim 0.4$) spin magnitudes, AGN feedback does not significantly impact the gas surrounding the BH, leading to frequent episodes of super-Eddington accretion: the BH spin only slowly increases, showcasing the ability of super-Eddington phases to delay the spin-up process. This becomes less efficient once the spin grows to higher values, as jetted feedback becomes more powerful: the frequency of super-Eddington episodes decreases, leading to a faster increase of $a$ from accretion at moderate sub-Eddington values. We discuss the question of predicting the BH spin when both super- and sub-Eddington regimes are involved in Sect. 4.3.

### 4.2. Evolution of the BH mass

In order to have a better understanding of the impact of the initial spin direction of the spin vector $\boldsymbol{a}_{\mathrm{ini}}$ on the super-Eddington mass evolution, we performed additional simulations, which all have $|a_{\mathrm{ini}}| = 0.7$ (as in Paper I) but different directions with respect to the large-scale gas distribution. We considered the following cases: in co-rotation with the galactic disc (sEdd_+z), in counter-rotation with the disc (sEdd_−z), and pointing into the disc (sEdd_y).

We show, in Fig. 3, the evolution of the spin parameter, $a$, (top panel) and BH mass, $M_{\mathrm{BH}}$, (bottom panel) for this set of simulations, as well as our fiducial simulation sEdd_0. For the same initial spin magnitudes, the BH is able to grow significantly more in mass when its spin is initially counter-rotating with the disc (sEdd_−z, solid green) than in the other configurations, as it takes ~15 Myr to align with the angular momentum of the accreted gas. As the realignment process occurs, the magnitude of the BH spin decreases down to $|a| \simeq 0.3$, while still counter-rotating. Since AGN jet efficiency in the super-Eddington regime is $\eta_{jet} \propto a^2$, feedback is weak. Super-Eddington episodes are frequent and thus able to contribute to effectively growing the BH mass.

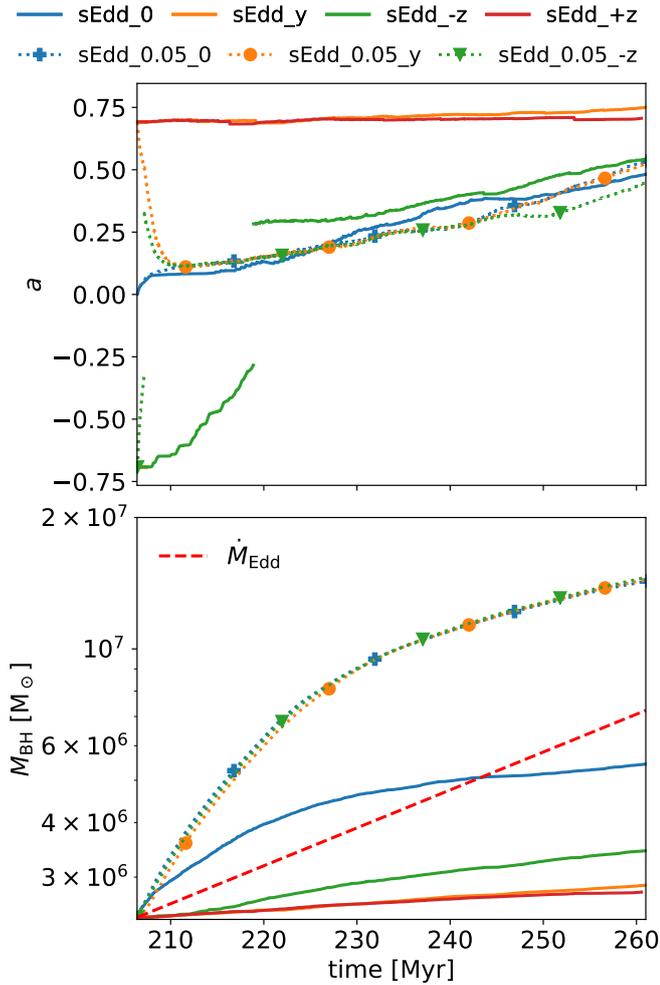

**Fig. 3.** Evolution of the spin parameter $a$ and BH mass $M_{BH}$ for different simulations. *Top*: evolution of the spin parameter $a$ for the sEdd_0 (solid blue), sEdd_y (solid orange), sEdd_-z (solid green), sEdd_+z (solid red), sEdd_0.05_0 (dotted blue), sEdd_0.05_y (dotted orange) and sEdd_0.05_-z (dotted green) simulations, from the moment super-Eddington was allowed ($t = 206.4$ Myr). *Bottom*: evolution of the BH mass $M_{BH}$ for the same simulations. For comparison is added the Eddington limit $M_{Edd}$ (dashed red). Different initial spin magnitudes and directions play a key role in the BH mass evolution, with slower BHs growing more. When the MADness factor $f_{MAD}$ is low (sEdd_0.05_0, sEdd_0.05_y and sEdd_0.05_-z), the growth in spin and mass are very similar amongst the simulations.

After the alignment with the disc angular momentum, the BH spin remains at moderate values ($|a| \simeq 0.3$–$0.4$) for ~20 Myr, allowing for a short period of efficient growth. At late times, the BH spins up to higher values, unleashing more and more powerful feedback during the increasingly rarer super-Eddington episodes, which leads to decreased BH growth rates.

Neither the BHs in sEdd_+z (solid red) or sEdd_y (solid orange) spend a lot of time in the super-Eddington regime, nor do they gain significant mass in comparison to their initial BH mass. Due to their persistently high spins, each super-Eddington event triggers powerful AGN feedback episodes and regulates BHs growth efficiently (see Paper I for discussion).

By comparing the BH of the sEdd_0 run (solid blue) with the BH from the other runs discussed in this section, it is apparent that a non-spinning BH is able to grow much faster and to higher masses than a (co- or counter-rotating) spinning BH of any other simulations presented this far. The BH that starts with $a_{ini} = 0$ is at the minimum of the AGN jet feedback efficiency in the super-Eddington regime – and also at a very low value of the radiative efficiency for the AGN quasar feedback which, although it is monotonically growing with the BH spin, is nearly flat with $\epsilon_r \simeq 0.05$–$0.1$ from $a \simeq [-1, 0.7]$ – until it sharply rises up to $\epsilon_r \simeq 0.4$ (see Zubovas & King 2019, for a different approach leading to similar conclusions for Eddington-limited BHs). As the BH spin slowly increases with accretion of gas, AGN feedback regulates the BH mass growth more efficiently, leading to an overall contribution to the BH mass growth dominated by the quasar mode at $t \simeq 240$ Myr (also see Fig. 2, bottom right panel).

We also compare the impact of the MADness parameter, $f_{MAD}$, on the BH mass and spin evolution, with three additional simulations, namely sEdd_0.05_0 (dotted blue), sEdd_0.05_y (dotted orange), and sEdd_0.05_-z (dotted green) in Fig. 3. The $f_{MAD}$ parameter has been decreased by a factor of 10, which corresponds to a jet feedback efficiency that is 100 times lower than the $f_{MAD} = 0.5$ simulations. For all three runs, BH mass and spin evolution are nearly identical, meaning that the BH spin initial direction and magnitude cease to have a significant impact on the BH evolution when a low MADness state of the accretion disc is assumed. We find that it only takes ≲5 Myr for the spin to align with the angular momentum of the disc. Because of the low MADness of the disc, the energy injected in the surroundings of the BH does not make a significant difference regarding the state of the gas near the compact object and gas can accrete on the BH almost unimpeded by the AGN feedback. Therefore, all BHs with $f_{MAD} = 0.05$ evolve in almost the same environment with the same orientation after the first 5 Myr of growth required for BH realignment.

In conclusion, both MADness and spin magnitude play a key role on the BH mass growth, as hinted in Paper I, and on the BH spin evolution. We find that the lower the MADness factor the more massive the BH is going to grow. For low MADness states ($f_{MAD} = 0.05$), for which the BH is not yet self-regulated, the growth of its mass and spin becomes independent of the initial BH spin. However, for high MADness states ($f_{MAD} = 0.5$), where BH is near self-regulation, then the lower the initial BH spin magnitude $|a|$, the greater the growth. The key lies in the AGN feedback, as for lower MADness and spins, feedback is weak and will not heat up the BH surroundings as effectively, allowing for more mass gained via super-Eddington accretion.

### 4.3. Predicting the spin evolution

The understanding gained with regard to the spin evolution in the super-Eddington regime from our simulations (detailed in the previous sections) leads to the question of whether it is possible to have an explicit expression of the spin, $a$, as a function of the mass accreted by the BH. We start out with some theoretical considerations, for instance: accretion from a thin disc of gas orbiting the BH led Bardeen (1970) to integrate the differential equation and obtain the evolution law (Eq. (3)). A similar approach can be used for a super-Eddington spin-up rate. Using Eq. (4), a maximally rotating BH gets spun-down by prograde accretion to $a \simeq a_{tr}$ after a moderate amount of accretion in super-Eddington of $M_{BH,fin} \simeq 2.64 M_{BH,ini}$. A maximally rotating BH gets spun-down by retrograde accretion to $a = 0$, after $M_{BH,fin} = 1.64 M_{BH,ini}$.

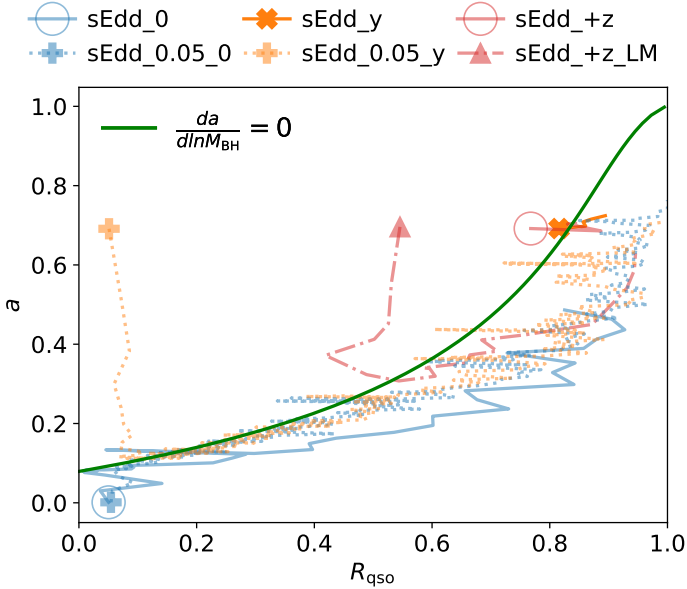

**Fig. 4.** Evolution of the spin parameter $a$ as a function of $R_{qso}$ for the sEdd_0 (solid blue), sEdd_y (solid orange), sEdd_+z (solid red), sEdd_0.05_0 (dotted blue), sEdd_0.05_y (dotted orange), and sEdd_+z_LM (dashed dotted red) simulations, from the moment super-Eddington was allowed. Equation (13) is shown in green. The markers show the starting $a$–$R_{qso}$ for each simulation. Each step represents a mass increase of $\Delta M = 8 \times 10^4\,M_\odot$. Only BHs in corotation ($a > 0$) during the entirety of the simulation are shown here. The sEdd_y and sEdd_+z BHs have self-regulated, therefore their trajectories on the $a$–$R_{qso}$ plane are very short compared to the rest of the simulations shown.

One can combine the quasar and super-Eddington contributions to the spin evolution of the BH. For the combined case, the spin variation of the BH for a given total mass variation of the BH $dM_{BH} = dM_{BH,qso} + dM_{BH,sEdd}$ is:

$$\frac{da}{d\ln M_{BH}} = (1 - R_{qso}) s_{sEdd} + R_{qso} \frac{s_{qso}}{\overline{E}_{qso}}, \qquad (12)$$

where we define $R_{qso} = dM_{BH,qso}/dM_{BH}$ to be the ratio of mass accreted in the quasar mode.

The ratio $R_{qso}$ and BH spin $a$ are dependent on one another: the spin variation depends directly on $R_{qso}$ and $a$; also, $R_{qso}$ also depends implicitly on $a$ because of how $a$ changes the feedback efficiency. Thus, the exact trajectories in the $a$–$R_{qso}$ plane are non-trivial. Nonetheless, Eq. (12) can guide us to understand how the spin should naturally evolve.

Let us first consider what would happen if a BH evolves at fixed $R_{qso}$. In this case, its equilibrium spin value can be obtained by setting $da/d\ln M_{BH} = 0$. We only consider spin parameters $a \in [a_{tr}, 0.998]$, as the alignment process is not taken into account in Eq. (12). Solving Eq. (12) for $da/d\ln M_{BH} = 0$ leads to the following relationship:

$$R_{qso}^{-1} = 1 - \frac{s_{qso}}{\overline{E}_{qso} s_{sEdd}}, \qquad (13)$$

which is shown as a green curve in Fig. 4. Any point on this line above $a_{tr}$ and below $a = 0.998$ represents an equilibrium value,

as contributions from super-Eddington phases (which decrease the spin magnitude) and quasar phases (which increase the spin magnitude) negate each other. Any BH not originally on the green curve would over time evolve towards it, and then be expected to stay at the point where it reaches the curve. For example, a BH always accreting at $R_{qso} = 0.5$ would evolve to $a \simeq 0.3$ from any initial spin value and then stay there.

As any point on the line represents an equilibrium value, and $R_{qso}$ is determined by the environment of the BH: a BH would not be expected to continue evolving in the $a$–$R_{qso}$ plane once it reaches the green curve, unless significant changes in the environment force a change in $R_{qso}$. Within the intricate environment of an evolving galaxy, the evolution in the $a$–$R_{qso}$ plane is more complex, as can be seen in Fig. 4, which shows the evolution of $a$ as a function of $R_{qso}$ for several simulations described in Table 1. Each step on the figure represents a mass increase of $\Delta M = 8 \times 10^4\,M_\odot$, and the markers show the $a$–$R_{qso}$ after the first $\Delta M$ has been accreted for each simulation.

We exclude simulations that display $a_{ini} < 0$ (i.e. sEdd_-z and sEdd_0.05_-z) from Fig. 4. Such simulations spin-up via gas accretion in the quasar and super-Eddington regimes, until they realign with the disc angular momentum and reach $a > 0$ (see top panel of Fig. 3), which is a process that is not included in Eq. (12). To avoid complications due to this omission, we only show simulations which have a positive spin parameter $a > 0$ for the entire duration of the run in Fig. 4. To investigate the role of $M_{BH}$, we add a simulation, sEdd_+z_LM, which has the same properties as sEdd_+z except that the initial $M_{BH}$ is half as massive.

The value of the ratio $R_{qso}$ depends on the gas properties (temperature, density, and gas velocity) in the BH accretion region, as well as the BH mass and the strength of the feedback. This is nicely demonstrated by the distribution of the initial coloured markers for all simulations in Fig. 4. We remind the reader that all[1] our simulations are variants of sEdd_0 and so, the initial properties of the accretion region are identical for all BHs when they are initialised in the simulation. For this reason, we could naively expect simulations with the same $a_{ini}$ to start their evolution at the same point in the $a$–$R_{qso}$ plane in Fig. 4. Instead, the range of $R_{qso}$ experienced by different BHs after a mass increase of only $\Delta M = 8 \times 10^4\,M_\odot$ (coloured markers) shows the dramatic impact of feedback on the spin evolution of BHs. The highest initial $R_{qso}$ (i.e. the orange cross and red circle markers) is found for sEdd_y and sEdd_+z at $R_{qso} \simeq 0.8$, while the initial $R_{qso}$ is lower for discs that are less MAD and BHs that have low spin values: sEdd_0.05_0, sEdd_0.05_y and sEdd_0 start with $R_{qso} \lesssim 0.1$. This is also true for lower BH masses, as sEdd_+z_LM has a ratio $R_{qso} \simeq 0.5$, since its feedback injection is in between the other cases already discussed (see Eqs. (7) and (8)). This confirms that the stronger the super-Eddington feedback, the higher $R_{qso}$ becomes, namely, the more mass is accreted at sub-Eddington rates, because super-Eddington episodes are limited by their own feedback. Weaker super-Eddington feedback occurs either with lower $\eta_{jet} \propto f_{MAD}^2 a^2$, thus lower MADness or spin magnitude, or with less massive BHs because of the smaller growth rates $\dot{M}_{BH} \propto M_{BH}^2$ (see Eq. (7)). In theory, $R_{qso}$ can also change with large-scale environmental effects, for example, a significantly cold inflow into the galactic centre might temporarily decrease $R_{qso}$, as the galaxy would feed the BH more efficiently than before. To have a better understanding of the continued evolution of BHs in the $a$–$R_{qso}$ plane, we separate the discussion of the BH spin evolution into three separate regions depending on the position of a BH in the $a - R_{qso}$ plane: above, on, and below the green curve.

---

[1] Apart from sEdd_+z_LM which has a less massive BH.

For BHs above the curve, $da/d\ln M_{BH} < 0$ (from Eq. (12)) as the super-Eddington regime dominates the BH spin evolution. As the BH gains mass, the spin will therefore decrease in magnitude, which is exactly what happens for sEdd_0.05_y and sEdd_+z_LM, until they reach the green curve. We note that these BH spins evolve at almost constant $R_{qso}$ until they reach the green curve, due to weak AGN feedback that does not significantly change the gas properties in the vicinity of the BH.

As mentioned before, if the BH lies on the green curve, $da/d\ln M_{BH} \simeq 0$ (see Eq. (13)) and if there are no changes to its environmental conditions, it will remain at fixed $R_{qso}$ and $a$. In the environment we have explored, super-Eddington phases become scarcer as the BH grows with time, leading to an increase of $R_{qso}$. The decrease in super-Eddington phases are caused by its powerful feedback, therefore when feedback is weak, $R_{qso}$ increases slowly and $da/d\ln M_{BH}$ remains small; consequently, the trajectory of the BH will remain close to the curve, but with a tendency towards larger $R_{qso}$. This is the case for the low MADness simulations (dotted lines): until the BHs have quadrupled their sizes (see Fig. 3), their spin magnitudes remain moderately small ($|a| \lesssim 0.3$) and due to their low MADness factors ($f_{MAD} = 0.05$), super-Eddington feedback is not powerful enough to disturb the BHs accretion region. Cases with low spin and low MADness, limiting the strength of super-Eddington feedback, are those that tend to remain closer to the conditions under which Eq. (13) was derived, and can therefore be described with this equation during this period of growth.

Finally, when the BH finds itself below the green curve, $da/d\ln M_{BH} > 0$ as spin-up due to accretion in the quasar mode is more important to changing the spin than the occasional spin-down during super-Eddington episodes. The quasar episodes tend to increase the spin magnitude, meaning subsequent super-Eddington episodes have a more powerful feedback which again increases $R_{qso}$ and moves the BH further away from the green curve. We find that after reaching the curve, all simulations shown in Fig. 4 eventually fall below the curve, and then stay below due to this mechanism. From this point onward, their trajectories are mostly driven by AGN feedback, which makes Eq. (13) insufficient for predicting spin evolution, since the effect of feedback cannot be included in a self-consistent way.

In conclusion, determining the evolution of $a$ for a BH that undergoes both quasar and super-Eddington regimes does not lend itself to a simple expression. However, we find that with some limits (e.g. for slowly spinning BHs), the evolution can be described by Eq. (13). At higher spins ($|a| \gtrsim 0.3$), the evolution is mostly driven by AGN feedback so the relationship in Eq. (13) no longer holds. In our work, in addition to sub-Eddington accretion, we find that coherent super-Eddington accretion coupled with relativistic jets, does not spin down the BH efficiently: all of our BHs stop experiencing super-Eddington accretion phases and, as a consequence, inevitably spin up to $a \simeq 0.998$.

For the analysis performed here, we used the super-Eddington spin-up parameter $s_{sEdd}$ from McKinney et al. (2012) (Eq. (4)). An alternative choice would have been to use the spin-up rate from Narayan et al. (2022) (Eq. (5)), shown in Fig. 1, which is more efficient at spinning the BH down for low spin values in comparison to McKinney et al. (2012). If we had used Narayan et al. (2022), the stable equilibrium curve we would find using $s_{sEdd} = s_{MAD,N22}$ in Eq. (13) would have shifted slightly below the green curve of Fig. 4, for all values of $a$. This would also lead to higher mass gained at a lower spin, that is, more massive BH in a similar amount of time. This would impact the spin evolution over long periods of time, as can be seen using the additional simulation with $s_{MAD,N22}$ discussed in Appendix A.

## 5. Discussion

An important constraint on models of SMBH growth is the observed mass-dependent spin distribution. Observationally, low-mass SMBHs tend to be rapidly spinning, suggesting growth via coherent accretion, whereas the more massive ones have modest-to-low spin magnitudes, indicating chaotic accretion and/or merger driven growth (e.g. Reynolds 2013, 2021; Soares & Nemmen 2020). In this paper, we study coherent super-Eddington accretion and ultimately find that regardless of the ability of super-Eddington feedback to spin down BHs, it only delays their spin-up by $\lesssim 100$ Myr. This is true for all initial BH spin magnitudes and direction probed here, so we predict that super-Eddington feedback will not have a significant impact on the global spin distribution.

Chaotic (or incoherent) accretion has been suggested to help BH seeds grow to $\gtrsim 10^8\ M_\odot$ (e.g. King et al. 2005, 2008; Zhang & Lu 2019; Zhang et al. 2020; Zubovas & King 2021). Despite the fact we limit our exploration to coherent accretion in this paper, we speculate that incoherent accretion may not lead to efficient super-Eddington spin-down if the BH is already rapidly spinning. For example, in Eddington-limited cosmological runs, spin-down induced by gas stripping from gas-rich mergers or counter-rotating filamentary accretion (Dubois et al. 2014b; Bustamante & Springel 2019) only decreases the spin magnitude to $|a| \sim 0.7$ (from $|a| \sim 1$), as the direction of the BH spin realigns with the gas angular momentum before the spin-down process is completed. If one of these events were to provide our BH with enough misaligned gas to reach accretion rates above the Eddington limit, the spin-down process would be less efficient than for the Eddington-limited BHs discussed before. The very powerful super-Eddington jets produced at very high spins for our BH would completely shut down the accretion of the misaligned material and would leave the BH spin almost unchanged ($|a| \sim 1$). In addition to this merger-driven gas misalignment effect, coalescence between maximally spinning BHs also leads to a decrease in the spin of the BH remnant (Rezzolla et al. 2008). While this effect can be dominant for setting up the spin of BHs in the passive and most massive galaxies (Dubois et al. 2014a), it is not clear how this should affect the spin evolution when super-Eddington accretion is included.

Our model for spin-up rate assumed that the MAD state obtained by McKinney et al. (2012; or Narayan et al. 2022), see Appendix A, which tries to capture the effect on BH growth of the uncertainties in spin-up rate) for sub-Eddington thick accretion discs can be extended to the super-Eddington regime. Although the geometry of the two discs are similar, their radiative efficiencies may differ significantly, and it remains to be tested through dedicated GRMHD simulations whether the spin-up rates in the super-Eddington regime is significantly different from the classical thick-disc state.

Episodes of accretion close to and above the Eddington limit have been found in cosmological simulations, however, the evolution of the BH spin under this critical regime has not been studied in this context. Whilst limited to an idealised setup of an isolated galaxy, the results obtained in this paper provide some insights regarding this problem. The delay of the spin-up that we found could occur for BHs in galaxies that are gas-rich, since they may have enough gas to trigger episodes of super-Eddington accretion. The BHs in gas-poor galaxies would likely have at most a few super-Eddington phases, which not therefore have great impact on the BH spin evolution. However, we cannot rule-out the possibility of a delayed spin-up in gas-poor galaxies, since these BHs would have more low-accretion sub-Eddington phases in the radio mode ($f_{\mathrm{Edd}} < 0.01$), which would spin down the BH. This is merely an educated guess, as in our simulations the radio mode did not contribute to growing the BH. Studying the BH spin evolution while assuming super-Eddington accretion in a cosmological context will be left to a future work.

## 6. Conclusions

In this study, we explore the evolution of the BH spin $a$ under the conditions of critical accretion above the Eddington limit in an isolated galaxy. The main focus here is to gain an understanding of the conjunction of a MAD in the super-Eddington regime, based on a thin disc undergoing accretion comparable to the limit, since these both vary the BH spin differently. Our main findings are as follows.

- The super-Eddington regime does not spin the BH down effectively, as $a$ inevitably increases towards ~0.998. It is only able to delay the certain spin-up of the BH from thin disc accretion. (Sect. 4.1)
- MADness and spin play a key role in BH growth: the lower the MADness the greater the growth. Similarly, the lower the spin magnitude, the greater the mass gained, thanks to weaker jets at lower spins. (Sect. 4.2)
- The combination of super-Eddington and quasar regimes does not lead to a simple expression for the spin evolution. Relatively low spins $|a| \lesssim 0.3$ are able to follow Eq. (13), but for higher spin magnitudes, AGN feedback makes it so that the evolution cannot be predicted. (Sect. 4.3)

*Acknowledgements.* We thank the referee for useful comments. RSB gratefully acknowledges funding from Newnham College, University of Cambridge and the ANR grant LYRICS (ANR-16-CE31-0011). This work was granted access to the HPC resources of CINES under the allocations A0120413449 made by GENCI. This work has made use of the Infinity Cluster hosted by Institut d'Astrophysique de Paris. We thank Stéphane Rouberol for smoothly running this cluster for us.

## Appendix A: Comparison of two different spin-up rates

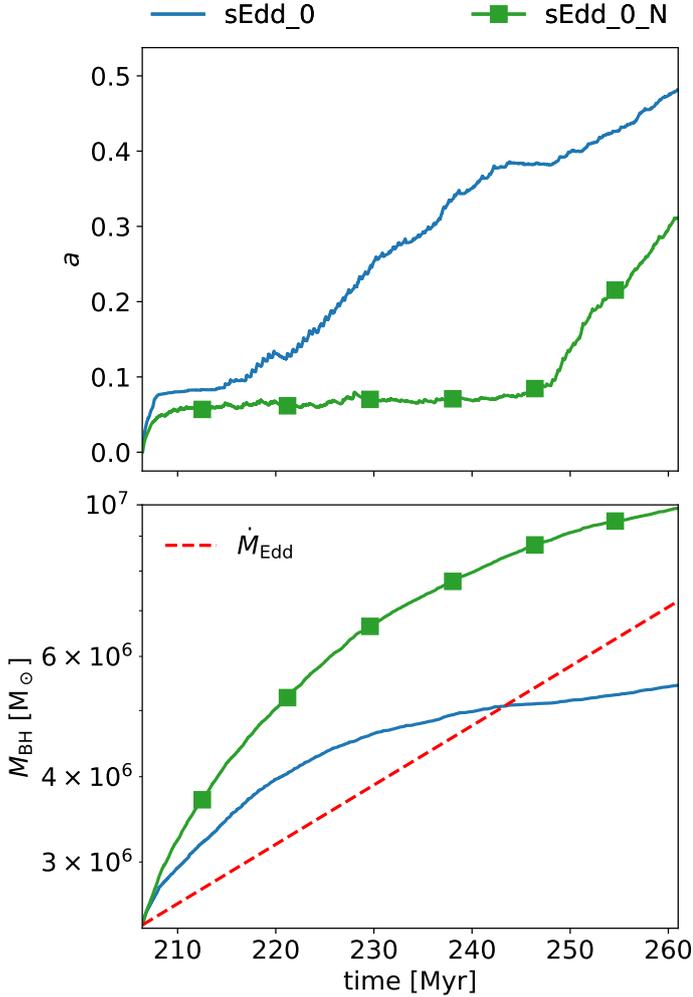

**Fig. A.1.** Evolution of the spin parameter $a$ and BH mass $M_{BH}$ for different simulations. *Top:* Evolution of the spin parameter $a$ for the sEdd_0 (solid blue) and sEdd_0_N (solid green) simulations, from the moment super-Eddington was allowed ($t$ = 206.4 Myr). *Bottom:* Evolution of the BH mass $M_{BH}$ for the same simulations. For comparison, the the Eddington limit, $\dot{M}_{Edd}$, has been added (dashed red).

In Section 2, we introduce different spin-up rates coming from McKinney et al. (2012) (Eq. 4) and Narayan et al. (2022) (Eq. 5) to compute the spin evolution of a BH producing jets. In the main body of the paper, we opt to use Eq. 4 from McKinney et al. (2012) in our isolated galaxy setup. Here, we investigate the differences between both spin-up rates, by performing a simulation similar to our fiducial run sEdd_0 (see Table 1), which only differs in the prescription of the spin-up parameter in the super-Eddington regime $s_{sEdd}$. For this run, named sEdd_0_N, we used the spin-up rate of Narayan et al. (2022) (see Eq. 5) and we compared the BH spin and mass evolution with our fiducial simulation in Fig. A.1.

The final mass of the BH reaches almost $10^7 \, M_\odot$ in the sEdd_0_N (solid green) simulation, twice as massive as sEdd_0 BH (solid blue). This faster and sustained growth is owed to much more frequent super-Eddington accretion episodes (lower $R_{qso}$), triggered more often due to the low spin magnitude that stays close to (a lower) $a_{tr} = 0.035$ (see Fig. 1, solid green) for close to 75 percent of its evolution, driven by the corresponding lower AGN feedback efficiencies. Because of this lower AGN feedback efficiency, it requires more mass for the BH to self-regulate, and to reach a state where $R_{qso}$ is sufficiently high to drive the BH spin up, and, hence, the rise of the BH spin is significantly delayed. Another way to see this is that the equilibrium curve $da/d \ln M_{BH} = 0$ is lowered in the $a - R_{qso}$ plane (Fig. 4), hence, producing lower spin values.